\documentclass[a4paper]{article}
\begin{document}

{\large

\begin{center}
QUANTUM PHYSICS AND REALITY\\[6mm]
{\bf Bernard d'Espagnat}\\[6mm]

{\it Laboratoire de Physique Th\'eorique}\\
{\it Universit\'e de Paris-Sud, 91405 Orsay, France.}
\end{center}
	
      Contrary to classical physics, which was strongly objective
      i.e. could be interpreted as a description of mind-independent
      reality, standard quantum mechanics (SQM) is only weakly
      objective, that is to say, its statements, though
      intersubjectively valid, still merely refer to operations of the
      mind. Essentially, in fact, they are predictive of
      observations. On the view that SQM is universal conventional
      realism is thereby refuted. It is shown however that this does
      not rule out a broader form of realism, called here `open
      realism', restoring the notion of mind-independent reality.
\vspace{5mm}
 
\noindent 1\,\,\,	KANT'S TWO TENETS
\vspace{5mm}

       Is reality something meaningful and is science steadily coming
       closer to a true depiction of it ? These two questions are
       obviously related and  belong very much to the domain of those
       to the analysis of which Evandro Agazzi most efficiently
       contributed. But I'll argue that they are distinct ones, and we
       shall see that at least one essential element in Agazzi's
       approach definitely points in this direction.

       On this subject the right way to proceed is obviously to begin
       by considering the second question, in which, of course, by
       `reality' what is meant is mind-independent reality, that is,
       the concept of something that would `be' even if conscious
       beings did not exist. Is science in principle capable of
       describing it ? As is well known, from the works of Kant, the
       positivists, the Marburg School neo-Kantians, etc. it became
       increasingly clear that a proof the answer is ``yes'' just simply
       cannot be given. On the other hand, the mere fact that a
       proposition cannot be proven is not a proof that it is
       wrong. It is therefore conceivable that science is able to
       reach at mind-independent reality - a position that may be
       termed conventional realism - and, a priori, the fact that the
       refinements brought by science to our primitive ideas about
       things make our knowledge more and more efficient may well
       induce us to think that way. But a priori it is also possible
       to consider that human finiteness will for ever prevent us from
       attaining such an ambitious goal and that what we are able to
       describe with certainty can only be (mathematically
       synthesized) relations between observed phenomena. 

      A fact that incited not only philosophers but also theoretically
      minded scientists to show a preference for the second standpoint
      was that, by successfully restricting the basic concepts to just
      a few ones - some of which, such as inertia, quite disconnected
      from direct experience -  and correspondingly emphasizing the
      role of laws, the Galilean revolution gave credit to the view
      that far from being a (tentative) description of the intrinsic
      properties of objects science might just be a construction
      grounded on prescriptive principles chosen a priori, in the
      sense of not being dictated by experience, and justified a
      posteriori through the fact that the observable consequences of
      the science derived from them agree with what is actually
      observed. Kant is generally considered to be, not really the
      originator but at least the main promoter of this idea, which
      indeed may be said to constitute one of the two fundamental
      tenets of his philosophy.  Which, of course, immediately raises
      the issue of how the said principles are to be chosen. And to
      this question Kant also gave quite a definite answer. So
      definite indeed that we may take it to be the second fundamental
      tenet of his thinking (in his writings the distinction between
      the two is hardly drawn but J.Petitot [1] pertinently stressed
      its rational necessity). It is the view that the principles in
      question are given a priori - that is, once and for all - by the
      very structure of our sensibility and understanding.

	Both tenets have a mentalist flavor. However, while mentalism
        is quite explicitly at the core of the second one, in the
        first one it is somewhat secondary. There, it merely
        corresponds to the fact that instead of being dictated by the
        ‘external world’ the prescriptive principles are chosen by us
        (with however the redoubtable final check that the overall
        theory they generate has to agree with observed data). But it
        implies nothing as to the nature of these principles so that a
        priori it may well be that those that work - that yield a
        theory matching with what is observed - happen to be
        compatible with an interpretation of the said theory as being
        a faithful description of mind-independent reality. 

	And indeed it is worth noting that this is more or less the
        direction in which the situation evolved during, roughly
        speaking, the classical era. For, in fact, the inferences Kant
        had drawn from his second tenet - such as the view that since
        we cannot but perceive objects as being embedded in a
        three-dimensional Euclidean space the physical space has to be
        three-dimensional and Euclidean - very soon became
        untenable. Followers of Kant were therefore obliged to
        progressively weaken the importance they imparted to this
        tenet - to reject the psychological foundation of Kantism, as
        is often said - and finally drop it altogether. Correlatively,
        as Michael Friedman showed in detail [2], while holding fast
        to the first tenet they repeatedly had to change its
        prescriptive principles, going from the three basic Newton's
        laws to Einstein's principle of equivalence, via Helmholtz'
        `principle of free mobility' and the special relativity
        principles. Such a sequence of changes, which has been termed
        ‘relativization of the a priori’, clearly would not have been
        possible if Kant's second tenet had been kept : the basic
        structures of human sensibility and understanding do not
        change so fast!
	Two points, in this, are worth noticing.

	 One is the fact that the expression ``a priori'' does not carry
         the same meaning in the two tenets. In the second one it has
         the original sense of something that cannot change since, by
         definition, it is given right at the start, once and for all,
         independently of our observing and theorizing. In the first
         tenet its meaning is much weaker and somewhat akin to that of
         a working hypothesis, even though a fully general, basic
         one. 

	The other point is that, as already observed, the first tenet,
        considered in isolation, is not, by nature, incompatible with
        the view that science reveals or will in the long run reveal
        what mind-independent reality truly is; for it may well happen
        that the successful prescriptive principles, though not
        consciously chosen with the purpose that they be compatible
        with the said view, still turn out to be so, perhaps because
        their progenitor was intuitively inclined to think that the
        view in question had to be true. 

	In other terms, a physicist of the early twenties, impressed
        by Einstein's achievements, could legitimately consider that
        the latter's fruitful method of deriving new physical concepts
        from pure mathematics had put him on the track of the truly
        good concepts, those by means of which the human mind may
        expect to, finally, describe mind-independent reality as it
        really is. And he could consider that on thinking this way he
        was not parting with what, in Kant's doctrine, remained
        unaltered and significant for research (what we called his
        `first tenet'). Anyhow it is a fact that mature Einstein was a
        realist himself and that the `prescriptive principles' (to use
        Kantian language) that he put forth are interpretable - and
        were quite unambiguously interpreted by him - as referring to
        mind-independent reality. 

	What all this seems to show is that, while the acceptance of
        Kant's above defined `second tenet' implies ipso facto the
        logical necessity of adhering to a mentalistic interpretation
        of physics, neither the Galilean revolution nor even (contrary
        to a widely held view) the acceptance of  the main lines of
        Kant's teaching entail the said necessity, provided that the
        acceptance in question be that, not of Kant's original
        doctrine but of a restriction of it to its `first tenet'
        which just means: to what it was bound anyhow to evolve to,
        due to the fact that after Kant's time the `prescriptive
        principles' of both geometry and physics had to be altered in
        ways that left  practically no place for psychological
        elements.

	This however is not to say that due to this evolution
        transcendentalism - meaning by this the general, philosophical
        conception put forward by Kant and developed as well as
        amended by his successors - merged into conventional
        realism. But in, say, the early nineteen-twenties the
        remaining difference between the two was essentially of an a
        priori and philosophical nature. It rested on the facts that
        theoretical physics is a mathematical synthesis of phenomena
        and that in transcendentalism it is generally considered an
        axiom that to interpret this synthesis as being, or as leading
        to, a faithful reconstruction of mind-independent reality
        would amount to attribute to the human mind powers exceeding
        its finiteness. Conventional realism obviously rejects the
        said axiom, so that clearly, in spite of the evolution
        transcendentalism had to undergo as we saw, between it and
        conventional realism a difference remained whose very nature -
        that of being a difference between two a priori conceptions of
        the capacities of the human mind - made it impossible that the
        question as to which doctrine is right be answered. 

	On the other hand it sometimes happens that by changing some
        elements in the debate science truly sheds light on
        philosophical problems. And we'll see in the next Section that
        such is the case with this one.
\vspace{5mm}

\noindent 2\,\,\,	QUANTUM PHYSIC'S `WEAK' OBJECTIVITY
\vspace{5mm}

	Up to the stage at which our schematic historical review led
        us, that is to the last and culminating achievements of
        classical physics, adepts of conventional realism could, as we
        saw, rationally entertain a strong belief not only in the
        existence but also in the accessibility of mind-independent
        reality. True, the legitimacy of this standpoint rested on the
        postulate - rejected by Kantians as above noted -  that  human
        mind is endowed with all the concepts that `fit' reality. But,
        after all, up to the advent of quantum mechanics this
        postulate seemed validated: Indeed, for a long time theories
        could be expressed just in terms of simple, fully intuitive
        concepts: space, position, motion, force etc. And we observed
        that later on - when these concepts proved insufficient -
        theories such as general relativity could successfully make
        use of similarly descriptive concepts, such as curved space,
        borrowed, this time, from mathematics. 

	On the advent of quantum mechanics it could naturally be hoped
        that it too would prove consistent with conventional
        realism. And indeed this hope was entertained, not really by
        the founding fathers of the theory, Bohr, Heisenberg
        etc. (even though the latter linked it with realism [3], but
        in another acceptation of that word) but, as it seems, by
        many of their successors. But was it fulfilled ? As long as we
        consider but static problems, such as calculating eigenvalues
        and the like, we may be tempted to answer ``yes'', for these
        problems are indeed qualitatively similar to classical
        ones. In some cases however what quantum mechanics yields is
        only the probability that, upon measurement of such and such a
        quantity, such and such value should be found (the `Born
        rule'). And at this place a discrepancy with conventional
        realism appears since when we say ``should be found'' we
        implicitly refer to somebody who finds, or to some detection
        device conceived of by human beings for the purpose of
        detecting. Admittedly, in everyday life the corresponding
        difficulty is easily removed. Instead of saying ``the
        probability is so and so that if we have a look at this card
        we'll see a King'' it suffices to say ``the probability is so
        and so that this card is a King'', thus attributing the need
        for probabilities probabilities just to lack of information. But this explanation calls for the existence of hidden variables and cannot therefore be transposed to standard quantum mechanics (SQM) that is, to the quantum formalism supplemented by the no hidden variable hypothesis. The point is that in such cases what SQM yields are just measures attached to the various possibilities. For imparting a meaning to them, hidden parameters being barred out, we have no other alternative than to resort to the facts that a measurement is a question and that a question cannot have several mutually incompatible answers. We {\it decide} therefore that the measures in question are probabilities of this {\it or} that result: an argument that involves the notion ``measurement'' in a very specific way, not detachable, as it seems, from voluntary action.

Admittedly, from the time of Louis de Broglie and David Bohm to the present one a number of brilliant physicists put forward theories that by introducing hidden variables could indeed reconcile conventional
        realism with the whole experimentally verifiable content of
        nonrelativistic quantum theory. Most interesting new vistas
        were opened thereby and indeed this line of research is still
        active and rightly so. It is fair to say however that up to
        now the efforts in this direction could no solve all
        difficulties, particularly concerning the matching with
        relativity theory, and that, consequently, no such theory
        could gain general acceptance.

	True, the (standard) quantum mechanical probabilistic
        statements can still be considered objective since they are
        valid for anybody. But clearly, contrary to those of classical
        physics (with due reservation concerning classical statistical
        physics) they are not expressed in terms a conventional
        realist would be happy with, since they involve us whereas in
        order to agree with conventional realism a statement should
        bear on what exists, not on what we see or intend to
        do. Statements of such a type I proposed to call ``weakly
        objective'' [4] in contrast with those, called ``strongly
        objective'', that a conventional realist can take to be bearing
        on mind-independent reality\footnote{Admittedly the fact that SQM is both weakly objective (i.e. intersubjective) and generally considered universal raises two conceptual problems. The first one is that, whereas classical physics could be developed without even mentioning mind, the fact in question, in view of the reference to ``us'' it implies, forces this vague, ill-defined notion into the picture. The second one is that minds are plural, which makes intersubjectivity something to be accounted for.

As a matter of fact, thinkers did not wait until the advent of quantum physics to start pondering over these problems. It is worth noting, however, that the first one did not worry them appreciably, the reason being that whoever does not entertain the predetermined idea that mind {\it must} be a product of matter has no reason to consider it to be one. As for the second one, note that intersubjectivity follows from the Born rule, so that in the eyes of whoever considers the Born rule an inherent part of SQM (or claim it can be derived from SQM's other axioms) intersubjectivity is accounted for by quantum theory. People reluctant to admit that much may still rest content on that matter by adopting Bitbol's view that : ``the point of support of science is not an explicit assertion concerning what other human beings see and think; it is simply a practice of communication which anticipates or presupposes the prefect interchangeability of positions amongst the members of the linguistic community.'' [5] 
}. While in theoretical works this
        particular difference between classical and standard quantum
        physics is quite often left implicit, personally I consider it
        an essential one since in my view it is the decisive fact that
        renders standard quantum mechanics incompatible with
        conventional realism.  
	Note that according to Kantism all scientific statements are
        weakly objective only. But within that doctrine they are so
        just because, in it, the finiteness of the human mind
        capacities has been posited a priori as an axiom. In standard
        quantum mechanics that same feature is an inherent element of
        the very structure of the theory so successfully used to
        account for observed data, an argument that, in my opinion and
        also, I think, in that of most scientists, has a greater
        weight.  
	As just noted, since the set of the quantum mechanical axioms
        includes the Born rule standard quantum mechanics is but
        weakly objective and this in turn implies that according to
        this theory conventional realism cannot hold true at atomic
        magnitude scale. Admittedly it could a priori be hoped that,
        at least, it would somehow hold true at the macroscopic level;
        and indeed when decoherence theory appeared, at first some of
        its promoters seemed to believe it could be interpreted that
        way. But it soon became clear that this is not really the
        case%
\footnote{In Zurek's 1982 seminal paper [7] on decoherence a remark
  seemed to mean that, just in virtue of decoherence, the instrument
  pointer already has, in every case and before we look, a definite
  position, period. This raised questions from such people as Elby
  [8], Healey [9], Bub [10] and myself [6], and later on Zurek
  explained that his statement had to be understood just in the sense
  that decoherence (`einselection') ``allows observers to anticipate
  what states in the Hilbert space have a `relatively objective
  existence' '' [11], relatively objective existence being defined
  operationally, by means of a criterion involving imagined
  measurement sequences.}. 
Mutatis mutandis, similar remarks could be made
        concerning Everett's relative state theory and many other
        valiant attempts at somehow reintroducing strong objectivity
        in the picture while keeping to the view that standard quantum
        theory is universal. Obviously this is not a proper place to
        review all the numerous valuable efforts that were made along
        this line throughout the whole century last:  a vast subject
        indeed, not susceptible of being summarized in a few
        pages. After having devoted much time and attention to its
        study (see Refs. [6] for a critical survey) what I may here
        confirm concerning it (and which all quantum theorists are
        aware of and is after all hardly surprising in view of the
        foregoing) is that, ingenious as all these interpretations
        are, all of them not only totally diverge from the commonsense
        picture of the world that we have in mind in our daily life%
\footnote{The search for such interpretations still actively goes
  on. In fact there are by now quite a number of them and, concerning
  my above mentioned survey, I have to grant that, extensive as it was
  meant to be, still some most interesting recent or fairly recent
  such proposals, such as those of Richard Healey [12], Michael Esfeld
  [13] and Carlo Rovelli [14] are not covered by it. Note however that
  all of them fall within the just described category in that none of
  them bears any resemblance with any of the various kinds of realism
  that laymen and philosophers normally have in mind when making use
  of that word.}
        but, what is more, are such that in them the link between what
        is claimed to 'really be' and what is observed is so strange,
        weak and indirect that its very existence gets extremely
        controversial. 

	Aware of the acuteness of the problem a number of conventional
        realists, remembering that no great theory is eternal, then
        considered the possibility that standard quantum theory should
        be replaced with a fully different one. Of course all its
        experimentally verified predictions had to be kept. But, as we
        know, along with them it also suggests something bizarre that
        normally is not observed, namely what we now call
        `entanglement-at-a-distance' [14], which, within a realist
        viewpoint, would imply that when two particles have once
        interacted they somehow remain linked together at a distance
        by some mysterious eternal bound.  Einstein, already, found
        this so-called nonseparability inacceptable. A priori the idea
        of a theory that would be free from such odd features seemed
        an attractive and realizable prospect. And it could be hoped
        that conventional realism would be recovered thereby.

	As we know these hopes were not substantiated. In 1964 John
        Bell [16] proved - independently of any specific theory - that
        to assume both conventional realism  and `locality' (i.e. to
        bar out any such things as the ``mysterious bound at a
        distance'') entails that some inequalities between measurable
        numbers must be satisfied whereas - he pointed out - they are
        violated by the quantum mechanical predictions. And indeed
        experiment confirmed this violation.  By now, barring most
        incredible `conspiracies' Bell's inequalities have  been
        experimentally shown to be violated in a wide range of
        correlation phenomena involving not only photons [17]  [18]
        [19] but also massive particles such as protons [20] and even
        systems such as heavy ions [21] and atomic ensembles [22]. It
        is true that practically all these experiments bore on systems
        conventionally labeled `microscopic'. However it is
        notoriously impossible to draw a sharp distinction between
        macroscopic  systems and microscopic systems endowed with mass
        so that, combining the just noted fact that nonseparability
        was observed on such systems with our knowledge that it is
        predicted by quantum mechanics - more and more considered to
        be a predictively universal theory%
\footnote{The nineteenth century progressive merging of magnetism with
  electrical science and optics with electromagnetism, followed in the
  twentieth century by the unification of electromagnetism and the
  weak nuclear force together withe the unified quantum treatment of
  this whole field and the strong nuclear forces is what convinced
  most physicists of the universality of the quantum laws (even though
  gravitation still remains outside the picture).} 
        - we are led to the, if
        not strictly inescapable at least extremely well grounded
        conclusion that any scientifically sound realism we may
        conceive of should be one in which the locality condition is
        violated. Which means that this realism must radically differ
        from the commonsense sense one we have intuitively in mind%
\footnote{The question might be asked whether Agazzi's  pertinent
  remark [23] that: ``reality is investigated by every science from
  its own point of view, consisting in the fact that only certain
  aspects of reality are considered'' makes it possible to refute the
  conclusion just arrived at. What, at first sight, speak for this
  possibility is the fact that sciences such as biology or neurology
  have laws of their own. However, it seems there can exist but one
  mind-independent reality and it is now well established that no
  science violates the laws of physics, so that what physics shows is
  impossible cannot be possible just because, within another science,
  other aspects of reality are considered. In other words, if  such a
  science displays an `aspect of reality' that is at variance with the
  conclusion we arrived at, a conventional realist will be forced to
  take that aspect - in a somewhat Kantian spirit - to be a mere
  phenomenon, that is, to merely be one of the ways mind-independent
  reality appears to all of us, due to our common sensorial and
  intellectual equipment.}. 

	It should of course be remembered that since the Bell theorem
        does not include the no hidden variable assumption within its
        premisses%
\footnote{Nor does it exclude it [24].}, 
considered in isolation the experimentally
        observed violation of the Bell inequalities does not yield an
        answer to the question, central to the present quest, whether
        or not mind-independent reality is knowable in principle (the
        non-relativistic hidden variable Broglie-Bohm theory is an
        example of a theory compatible with the answer ``yes it
        is''). But the said violation yields nevertheless a most
        important clue in this respect  since it shows that even if,
        in the hope of recovering conventional realism, we dropped the
        no hidden variable assumption this would hardly help us since
        whatever description we could construct of mind-independent
        reality would anyhow basically differ from any image we try to
        build up of it%
\footnote{Note that this is quite compatible with Agazzi's insistence
  on the fact that concerning the notion of reality no presupposition
  should be made bearing on its nature, ``the only requirement we need
  for reality being therefore that of existing, and not that of
  existing as a reality of such-and-such a kind'' [23].}.

	Such a state of affairs may remind us of structural realism
        that is, the idea, already present in Poincar\'e's works [25],
        that we know, not the objects themselves but only the
        relationships between them (the equations), one implication of
        which is that the monadic propositions contemporary physics
        makes use of when dealing with electrons, quarks and so on
        merely describe phenomena in the Kantian sense of the
        word.True, structural realism also implies that nevertheless
        we can know something of mind-independent reality, namely its
        structures, and this is an interesting idea. But independently
        of it the foregoing holds true. Consequently structural
        realism agrees with standard quantum theory in stating that
        electrons, quarks, atoms etc. and their attributes are not
        elements of mind-independent reality. To repeat, they are just
        phenomena, and constitute what may be called empirical
        reality.

	To all this it should be added that under extremely general
        assumptions decoherence theory nicely accounts for the
        appearance (to us) of a classical world (with distinct
        macroscopic objects endowed with definite, robust attributes
        and so on), thus extending to macroscopic objects the above
        stated conclusion… and sparing us the inconvenience of having
        to resort to the esoteric (though thinkable) idea that this
        classical world exists in an absolute sense even though the
        parts it is composed of do not enjoy such a status. 

	To this impressive list of scientific arguments speaking
        against conventional realism it is perhaps appropriate to add
        the well known ``no God's eye view'' one, a philosophical
        argument but still one often put forward by scientists. It
        consists in the observation that being ourselves parts of the
        world it is quite natural indeed that far from being a
        strongly objective picture the view we get of it should be
        highly dependent of our position within it, same as the view
        we have of a gulf when sitting on one of its beaches is
        altogether different from the (correct) one we should get of
        it were we on a plane or an orbiting satellite.

	Considering that quantum theory is by now generally considered
        to be universally valid what all this shows is that concerning
        the question whether or not mind-independent reality is, in
        principle, scientifically knowable the advent of standard
        quantum mechanics radically turned the tables. For, as we saw,
        while at the end of the classical era the view that physical
        knowledge is, in essence, a description of mind-independent
        reality (even though its acquisition rests on freely imagining
        laws) seemed quite a legitimate and natural one, standard
        quantum mechanics is quite definitely uninterpretable that way
        since in it objectivity is but weak. In other terms, while
        both commonsense and classical physics strongly indicated that
        only the part of Kant's doctrine compatible with a
        naturalistic conception of knowledge -  the part defined above
        as being its first tenet - had to be kept, nowadays if we want
        to make our general philosophical conceptions consistent with
        our confidence in contemporary physics - predominantly
        anchored on the basic axioms of standard quantum mechanics -
        we must give up this naturalistic interpretation
        altogether. This means that when Kant put forward an
        essentially mentalist conception of scientific knowledge he
        was basically on the right track.

	Now does this imply he was right in putting forward his
        `Copernician revolution'? The answer, of course, is “yes” if
        we identify the latter, as G.Boniolo did, with the axiom that
        ``we are able to know, also perceptively, only what we have a
        theorization about'' [26], for such a view is nothing else than
        the above ’first tenet’. But the answer is ``no'' if if we take
        this revolution to have the more precise meaning: ``while it
        remains true that the basic elements of physical knowledge are
        to be described, such descriptions must be given in terms of
        the a priori concepts dictated by the very structure of our
        sensibility and understanding'', for then we cannot escape the
        pitfall that the said a priori concepts include Euclidean
        space and similar notions. In other words, while, as we saw,
        contemporary scientific knowledge can only be of a mentalist
        form (at least in physics), the way of being so Kant ascribed
        to it must be rejected. Fortunately standard quantum mechanics
        yields here a working answer for besides being weakly
        objective the probabilistic statements linked with the Born
        rule have another significant feature not obviously
        generalizable to all weakly objective statements: They are
        explicitly predictive of observational results. And it is a
        striking fact indeed that while, viewed as just rules for
        predicting future measurement outcomes the quantum mechanical
        axioms (with their apex, the Born Rule, included) were
        immensely fruitful in explaining and predicting effects and
        never led to erroneous predictions, all attempts at
        interpreting them as descriptive laws led, up to now, to hosts
        of consistency difficulties. This shows, I think, that quantum
        mechanics is in fact an essentially operationalist  theory and
        that, as such, it does not describe anything - not even mere
        phenomena in the Kantian sense - except in a sort of
        figurative and somewhat allegoric way, grounded on
        abstractions being made from contexts that, if questions other
        than the one at hand were considered, could not be
        legitimately abstracted from.

	But do these inferences, that I deem conclusive concerning
        physics and even science in general, imply that the notion of
        a mind-independent reality is altogether meaningless ? In the
        next sections I shall try to explain why I consider they do
        not.
\vspace{5mm}

\noindent 3\,\,\,  ARGUMENTS INDICATING THAT THE MIND-INDEPENDENT
REALITY NOTION HAS TO BE KEPT
\vspace{5mm}

	When Kant put forward his conception he declared the
        ‘thing-in-itself’ (another name for mind-independent reality)
        to be unknowable; but he carefully kept the notion. Later the
        neo-Kantians discarded it, but there are reasons for not
        following them on this issue. True, none of these reasons
        emerges from science. They all are but of a philosophical
        nature. However, I take the three following ones to be well
        worth considering.
\vspace{5mm}

\noindent Reason 1\\
	This is just the fact that we are not permitted to believe in
        any theory we like. It recurrently happens that extremely
        ingenious theories, quite elegant mathematically, are
        disproved by experiments. Clearly something resists us and,
        though not impossible logically, the idea that also this
        something should be us seems in fact hardly conceivable. In
        other words there must exist something that is not
        mind-dependent. Scientists being deeply aware that their
        theories have to pass the test of experiment, this reason is
        presumably the one that will sound the most convincing to
        them.
\vspace{5mm}

\noindent Reason 2\\
	As many thinkers pointed out, Descartes' Cogito does not prove
        the existence of a personal `I'. Nor does it prove that
        thought does not derive from matter. And it is not a
        syllogism. But still, it expresses our inner experience of the
        existence of something - consciousness, thought - existing not
        just as a representation (for consciousness would then be a
        representation to itself, which makes no sense) but as
        something - the seat of representations - definitely more
        fundamental than any representation : in other words existing
        on a deeper mode. As we see, this inner experience we have
        suffices to show that such a notion of existing in a deeper
        mode than phenomena do, a mode we may agree to call `in se',
        has a meaning. And, since to think one must first exist - and
        precisely in that sense -  it is clear that such a basic
        notion of existence is prior to that of thought. Now, when
        things are considered under this light the idea that existence
        in se is not restricted to consciousness appears to be, if not
        compelling, at least extremely natural. Be it only because it
        would seem quite presumptuous indeed to believe that the whole
        of what exists in such a strong sense is, in essence, us.
\vspace{5mm}

\noindent Reason 3\\
	It is just the remark that at least in its most commonly
        received version [27]%
\footnote{The hypothesis has also been considered [28] that only
  structures are real, but there are objections to this [29], one of
  them (elementary but still not easily disposed of) being that
  relations without relata are unthinkable.} 
structural realism includes the
        assumption both that mind-independent reality exists and that
        most important elements of it - the objects - will for ever
        remain unknowable. Similarly one of the above mentioned
        objections to conventional realism, namely the `no God's eye
        view' one, while showing that mind-independent reality is not
        knowable or at least, not ‘as it really is’, still obviously
        assumes that this reality exists. Consequently whoever grounds
        his/her rejection of conventional realism on either structural
        realism or the `no God's eye view' argument cannot
        consistently deny the existence of the unknowable. He/she has
        to take the notion of an unknowable reality to be a meaningful
        one (and the same, of course, is true concerning the
        cosmologists who take the `multiverses' notion into
        consideration).
\vspace{5mm}

	A `Reason 4' - considered very strong by the persons in the
        opinion of whom it is scientifically established that
        consciousness emerges from matter - is just that, obviously,
        neither transcendentalism nor the conceptions that, similarly,
        define whatever is meaningful by referring to sensations,
        perceptions etc., that is, ultimately, to consciousness, are
        logically compatible with the opinion in question. Personally
        I do not share the said opinion so that I do not take this
        fourth reason to be a valid one. Still, I believe it is one
        that has had a prominent role in inciting physicists to look
        for ways of recovering realism.
 \vspace{5mm}

	I think this state of affairs justifies a renewed attempt at
        philosophically reflecting about realism even though it is
        clear from the content of Section 2 that anyhow, whatever
        realism we may concoct will have to differ very much from
        everything we think we learned from both experience and
        experiment (as H.Zwirn put it: ``it seems that, for the first
        time in the history of philosophy, choosing to believe in the
        existence of a world external to all observers and roughly
        consonant with what we perceive  is not possible any more''
        [30]). In the following Section such an endeavor is
        described%
\footnote{The idea that consciousness and empirical reality (the
  phenomena composing the world) generate one another is to be found
  in the works of several authors including myself [31] and is
  sometimes considered a substitute to the mind-independent reality
  notion. This view is questionable however for to claim the phenomena
  are generated by consciousness means they are appearances to it and
  it seems unthinkable that a mere appearance should produce - or
  generate - anything, consciousness in particular. And resorting to
  the well-known alternative definition of causality, namely the
  `regularity' one, according to which a causal link is just an
  instantiation of some general law, does not help since no known law
  exists that could here be referred to.}. 
\vspace{5mm}

\noindent 4\,\,\,	TWO NOTIONS OF REALITY, OPEN REALISM
\vspace{5mm}

	 In philosophy realism is usually defined to be the idea that
         the notion of a mind-independent reality is not void (there
         truly is such a reality)  and  that the said reality is
         knowable by us, at least in principle. In essence this is the
         conception Evandro Agazzi referred to when he described :
         ``the basic `realist' attitude of any knowing subject'' to be
         that ``this subject […]  believes [his] knowledge actually to
         be of something other [than himself]'', and stressed a few
         lines further that ``this ontologically distinct world must
         nevertheless be accessible to the subject'' [32].

	Obviously, according to this definition - and these Agazzi
        quotations as well - conventional realism is a combination of
        two distinct views: the idea that the notion of a
        mind-independent reality is meaningful and the one that this
        reality is in principle knowable by human beings. Now, it is a
        fact that from a purely logical point of view nothing forces
        us to combine the two together. In other terms, the notion of
        a reality being what it is quite independently of whether or
        not it is partly or totally knowable is by no means logically
        inconsistent. True, such a statement may at first sight sound
        surprising considering the classic objection that the idea of
        an absolutely (alias ‘in principle’) unknowable reality seems
        meaningless. A moment attention shows however that the
        inference: ``if an idea refers to something unknowable it has
        no meaning'' is consistent only if the meaning  of the phrase 
        ``to have a meaning'' - or ``to have no meaning'' - is itself
        defined by (partly or totally) referring to us (to our
        abilities at perceiving or conceptualizing or acting or
        etc.). But this ultimate reference to us - a characteristic
        feature of idealism - is precisely what realism is meant to
        avoid. In other words, as soon as we have endorsed the notion
        of a mind-independent reality we have to consider it to be a
        primary one that is, one that does not have to be defined or,
        in other terms, has a meaning by itself, quite independently
        of  our abilities. And this invalidates the objection%
\footnote{In other terms, pertinent as the celebrated Wittgenstein
  axiom ``whereof we cannot speak, thereof we must keep silent'' may be,
  it remains true that even though such an unknowable reality is not
  something we should try to describe, still we may entertain the idea
  of its existence.}. 

	Consequently, it is appropriate to take into consideration two
        very different versions of realism, both meaningful. From now
        on we shall, by convention, call `open realism' just the idea
        that the notion of mind-independent reality has a meaning and
        is valid and we shall keep the expression `conventional
        realism' for designating, as above, open realism supplemented
        with the assumption that the entities accepted scientific
        theories deal with, far from being mere appearances, are
        elements of mind-independent reality. The open realism notion
        may, at first sight, look disconcerting but, as already noted,
        in fact it is, in a way, implicitly postulated by the
        structural realists as well as by the many philosophers and
        scientists who quite pertinently point out that since we
        ourselves are parts of reality it would be most preposterous
        on our part to maintain we can get a ‘God's eye view’ of it.

	Of course, for open realism not to be a purely gratuitous
        conjecture devoid of any bearing on what we think or feel it
        must be assumed that between the reality it refers to and
        ourselves some causal links somehow exist. In other words it
        must be assumed that the said reality generates in our mind
        some pieces of knowledge.  Considered alone, this hypothesis
        in no way implies that we are thereby informed about what
        reality `truly is', for the said pieces of knowledge may well
        be mere recipes for predicting some of our future impressions
        (or they may even be some pieces of information about it, but
        uncertain, and so scarce as not to allow for a
        description). Hence the hypothesis in question, contrary to
        the one of conventional realism, lays itself open to no sharp
        refutation either from a scientific or a philosophical
        provenance. 

	Things being so, I proposed [31,6,4] a balanced view
        consisting in considering that the mind-independent reality
        notion is meaningful, that this entity - which most presumably
        is not imbedded in space-time - truly 'is' (exists, and not
        merely in our thought), but is essentially unknowable (in the
        sense of not being describable with concepts), so that what
        both commonsense and science refer to and are able to really
        describe is merely what it generates in our mind and we called
        empirical reality. Obviously this distinction between two
        notions of reality is consonant with quite a fundamental one
        the need for which Evandro Agazzi powerfully stressed, to wit
        the distinction between the sense and the reference of a
        concept (see e.g. [23]). As I see things, its necessity rests
        on two pillars.

 	The first one is the circumstance that science, and in
        particular physics - just due to the continuous enlargement of
        its field of research (toward the very small, the very large
        etc.) - increasingly meets with problems traditionally
        considered to lie in the realm of metaphysics, so that it has
        to be more and more careful to avoid admixture of some
        imprecise notions to be found in the latter, which implies
        that to deserve the qualification ‘scientific’ an inference
        must be grounded on references fulfilling stricter and
        stricter requirements. Since, considering the arguments of
        Section 2, it seems fairly clear that contemporary physics
        essentially is a purely operationalist theory the said
        references must ultimately be to intersubjectively valid
        elementary perceptions (`perception' meaning here the mental
        act of perceiving). 

	And the second pillar is just the set of reasons enumerated in
        Section 3 (see especially Reason 2), which, I think,
        convincingly indicate that the said human elementary
        perceptions simply cannot constitute the whole of what
        exists. Which means that the intersubjective mental events -
        which are, in the last resort, all that physics may refer to
        and report on - do not include everything that is
        existentially meaningful. As above noted, these reasons are
        not scientific ones; but  the fact that conventional realism
        failed is scientific (i.e. is grounded on arguments of a
        scientific nature), so that, even for a scientist, the choice
        between possible rational standpoints is restricted. Roughly
        speaking it is between the one described here and pure
        transcendentalism and the arguments of Section 3 speak in
        favor of the former.

	They do so all the more as they seem to indicate that, as
        noted above, in some way mind-independent reality produces in
        our minds elements of knowledge, although ones of such a
        nature as not to allow for actual descriptions of it. Of
        course the most powerful evidence that this is indeed the case
        is just precisely that physics makes predictions that turn out
        right: for, to repeat, such predictions follow from rules
        (ultimately, the basic quantum mechanical axioms) that
        unquestionably are elements of knowledge, are most unlikely to
        be mere inventions of ours (see Section 3, Reason 1) and must
        therefore proceed from something that is not ‘us’. Now, once
        this has been duly acknowledged it becomes possible and even
        rather natural to hold that this ‘something’ also imparts to
        us%
\footnote{`Us' who may be, or not be, just ‘pieces’ of it… Such purely
  metaphysical questions are fascinating but, unfortunately, seem to
  lie somewhat beyond the borderline of the domain human thinking may
  safely venture into.} 
other elements of knowledge, some of which, conceivably,
        may even indicate elements of itself although we cannot know
        whether or not this is the case (the fundamental constants,
        for example, fall into that category). Finally, since we
        cannot know anything for sure concerning this mind-independent
        reality - this `ground of things' - it is also conceivable -
        even though merely conceivable ! - that some elements of our
        experience are dim signs of its true nature%
\footnote{M.Bitbol noted [33] the similarity between this hypothesis
  of mine and the one Russell put forward in {\it The Analysis of Matter},
  according to which without having any knowledge of the
  thing-in-itself, still, we enjoy an indescribable  acquaintance with
  it.}. 
So that,
        contrary to what was the case in the classical physics
        glorious days, admirers, say, of classical music, or nature
        lovers, may quite consistently and rationally choose to
        entertain the belief that the emotions they feel when
        listening to a concert or just looking at appealing landscapes
        do open to them some kind of a window looking out on the
        `ground of things'.

	Admittedly this conception raises new questions concerning
        both matter and consciousness. Regarding matter a received
        view during the said classical days was that it is, by
        definition, what the physical science deals with, which
        implies it was considered knowable, at least in principle. And
        when, even today, we make use of the word ‘matter’ we have in
        mind the idea of something the most important features of
        which scientific research is somehow able to master. Clearly
        therefore, short of imparting to it a meaning opposite to the
        one it has in common language we cannot use it for designating
        the essentially unknowable mind-independent reality we formed
        here the notion of. In common language matter means something
        like the set of the atoms or elementary particles or
        etc. (together with the corresponding fields) and this, in the
        conception in hand, implies viewing matter just as a set of
        phenomena, that is, a set of mere appearances to
        consciousness. 

	Concerning consciousness this, in turn, has the immediate but
        apparently baffling consequence that it renders meaningless
        the hypothesis considered by most scientists to be the most
        reasonable one, namely that consciousness emerges from matter
        : for indeed in the present context this hypothesis would
        amount to asserting that consciousness emerges from
        appearances to consciousness,… which makes no sense. Bearing
        this in mind, an idea that seems rather natural is to
        substitute to the said hypothesis the one that consciousness
        emerges from mind-independent reality: a view that while
        sounding rather similar to the hypothesis in question in fact
        altogether differs from it since it means consciousness
        emerges from something that lies beyond our intellectual
        grasp. 
\vspace{5mm}

\noindent REFERENCES\\
1  - 	J. Petitot, Objectivit\'e faible et Philosophie transcendantale,
in {\it Physique et R\'ealit\'e},  M. Bitbol and  S. Laugier eds, Editions
Fronti\`ere, Paris (1997).\\
2  - 	M. Friedman, `Einstein, Kant and the Relativized A Priori', in
{\it Constituting Objectivity}, M. Bitbol, P. Kerszberg and J. Petitot
eds, The Western Ontario Series in the Philosophy of 	Science 74,
Springer (2009).\\
3  -	W. Heisenberg, {\it Physics and Philosophy}, Harper \& Brothers, New
York (1958).\\  
4  - 	B. d'Espagnat, {\it Conceptual Foundations of Quantum Mechanics},
W. A. Benjamin,  Reading, MA (1991), 2d ed reset (1976), 4th
ed. Perseus Books, Reading MA (1999).\\
5 - M. Bitbol, ``The problems of other minds: a debate between Schr\"odinger 
and Carnap'', {\it Phenomenology and the Cognitive Science}, 3 (1), 115-123, 2004.\\
6 - B. d'Espagnat, {\it Veiled Reality}, Addison-Wesley 1995, 2d
ed. Westview Press, Perseus 	Books 2003; {\it On Physics and Philosophy},
Princeton University Press, (2002).\\
7  - 	W. H. Zurek, Phys. Rev. D 26, 1862-1880 (1982).\\
8  -	 A. Elby, `Decoherence and Zurek's existential interpretation
of quantum mechanics', in {\it Symp. on foundations of modern
physics}, P. Busch, P. Lahti \& P. Mittelstaedt eds (1993).\\
9  -	 R. Healey in {\it Quantum measurement, Decoherence and lodal
interpretations}, G. Hellman \& 	R. Healey eds., Minessota Studies in
the Philosophy of Science, (1998).\\
10 - 	J. Bub, {\it Interpreting the quantum world}, Cambridge University
Press (1997).\\
11  - 	W. H. Zurek, `Decoherence, einselection and the existential
interpretation (the rough guide)', in {\it Phil. Trans. R. Soc. Lond.} A 356, 1793-1821 (1998).\\
12 - 	R. Healey, {\it The Journal of Philosophy}, 88, 393-421 (1881).\\
13  - 	M. Esfeld - {\it Holism in Philosophy of Mind and Philosophy of
Physics}, Kluwer (2001).\\
14  - 	C. Rovelli, {\it International Journal of Theoretical Physics}, 35,
1637-1657), World Scientific 	(1996).\\
15 -	E. Schr\"odinger, `Discussion of probability relations between
separated systems', {\it Proc. 	Camb. Phil Soc.} 31, 555-563 (1935).\\
16  - 	J. Bell, {\it Physics}, 1, 195-200 (1964).\\
17  - 	S. J. Freedman and J. F. Clauser, {\it Phys. Rev. Lett}. 
28, 938 (1972).\\
18  - A. Aspect, J. Dalibard, and G. Roger, {\it Phys. Rev. Lett}. 
49, 1804 (1982).\\
19  - G. Weihs et. al, {\it Phys. Rev. Lett}. 81, 5039 (1998).\\
20  - M. Lamehi-Rachti and W. Mittig, {\it Phys. Rev.} D 14, 2543(1976).\\
21  - M. A. Rowe et al., {\it Nature} 409, 791 (2001).\\
22 - 	D. N. Matsukevich et al., {\it Phys. Rev. Lett}. 96, 030405 (2006);
C.-W. Chou, et al., {\it Science} 316, 1316 (2007).\\
23 - 	E. Agazzi, {\it Realism and Quantum physics}, (Introduction and
article), E. Agazzi ed., Poznan 	Studies in the Philosophy of the
Sciences and the Humanities, Rodopi, Amsterdam (1997).\\
24 -	J. S. Bell, `La nouvelle cuisine', in {\it Between science and
technology}, A. Sarlemijn \& P. Kroes eds., Elsevier/North Holland,
97-115 (1990).\\
25 - 	H. Poincar\'e, {\it La science et l'hypoth\`ese,} 
Flammarion, Paris (1902).\\
26 - 	G. Boniolo, `What does it mean to observe physical reality ?',
in {\it The reality of the unobservable,} 
E.  Agazzi \& M. Pauri eds., Kluver,
Dordrecht (2000).\\
27 - 	J. Worrall, `Structural Realism, the Best of Both Worlds?',
{\it Dialectica} 43, 99-124 (1989).\\
28 - 	J. Ladyman, `What is structural realism ?', in {\it Studies in the
History and Philosophy of Science,} 29,409-424 ((1998).\\
29  - 	M. Bitbol, {\it De l'int\'erieur du monde,} Flammarion, Paris (2010).\\
30  -  H. Zwirn, {\it Les limites de la connaissance},  Odile Jacob, Paris (2000).\\
31  -  B. d'Espagnat - {\it In Search of Reality,} Springer New York, (1983).\\
32  - 	E. Agazzi, `Observability and referentiability', in {\it The reality
of the Unobservable,} E. Agazzi 	\& M. Pauri eds., Kluver,
Dordrecht (2000).\\
33 - 	M. Bitbol, `Troisi\`eme dialogue avec Bernard d'Espagnat', in
{\it Philosophie de la physique,} L. Soler ed., L'Harmattan, Paris
(2006).

}
\end{document}